# A Simulation Study of Electric Field Engineering with Multi-Level Pinned Photodiodes for Fast and Complete Charge Transfer

Hamzeh Alaibakhsh, and Mohammad Azim Karami

*Abstract*—In a CMOS image sensor pixel, fast and complete charge transfer from pinned photodiode (PPD) is desired and necessary in some applications. In special cases such as time-of-flight imaging or large pinned photodiodes, the PPD potential well shape highly affects the charge transfer performance and should be engineered carefully. In the present work, a PPD structure named multi-level PPD is introduced and examined through simulation study. Moreover, a fast and effective way to analyze the pinning process for a lag-free design is introduced. It is concluded that the proposed PPD achieves fast and complete charge transfer without additional implementation masks or process steps. The proposed PPD is compared with a similar conventional rectangular pixel and 31% reduction in the charge transfer time is observed.

*Index Terms*— Pinned Photodiode; Lag-Free; Charge Transfer; CMOS Image Sensor; Charge-Coupled Device

## I. INTRODUCTION

PINNED photodiode (PPD) is responsible for converting light to electrons and is the main building element in most of charge-coupled devices (CCD) and complementary metal oxide semiconductor (CMOS) image sensors (CIS) [1-4].

The PPD doping profile and structural design for superior charge transfer performance is widely practiced [5-9]. In previous works, the focus was directly on the electric field while the importance of pinning process was ignored. In present work, the electric field engineering is performed with focus on pinning process and a new PPD structure is proposed. The new structure is conceptually similar to [9] though free from additional mask needed for extra implantation process in [9]. Moreover, there is no need for additional thermal diffusion steps used in [9] to smooth out the doping profile. In the present work, electric field engineering can be performed with minimal fabrication process modifications (only by PPD mask modification) with the focus on pinning process.

Corresponding Author: Mohammad Azim Karami (karami@iust.ac.ir).
H. Alaibakhsh and M. A. Karami are with School of Electrical Engineering, Iran University of Science and Technology, Tehran 1684613114, Iran (e-mail: h_alaibakhsh@elec.iust.ac.ir; karami@iust.ac.ir).

Understanding and characterizing the pinning process is essential in the PPD design process. A pinning process measurement method is proposed in [10] and is further described in [11]. Moreover, detailed mathematical descriptions of the pinning process can be found in [12]. In the present work, a fast and effective simulation approach is proposed to characterize the pinning process based on the measurement scenario of [10].

The proposed approach is used to characterize a standard PPD and also two new PPD structures. The new structure is successful in forming considerable electric field to guide electrons towards the transfer gate (TG) and hence, increase the charge transfer rate.

Section II present the pinning simulation method. In section III, new PPD structures are proposed. Section IV presents the results and discussion and section V concludes the paper.

## II. PINNING SIMULATION METHOD

In this section, the pinning simulation method (PSM) is presented and used in the analysis of a conventional 3-D PPD CIS with a rectangular PPD [13]. The simulations are performed with a commercially available software in 3D [14, 15]. Similar to the injection phase of the measurement method of [10], transfer and reset gates are biased with the ON potentials of 3.3 V. Then, the reset drain potential is swept from 0 to 3.3 V. The difference between the proposed simulation method and the measurement method of [10] is that in these sets of simulations, the potentials and quasi-fermi level (QFL) energies of the PPD and the floating diffusion (FD) and also the PPD electron population can be probed. Hence, there is no need to do the charge transfer step after the injection phase and more importantly, there is no need to repeat the process for each reset drain potential step.

Another important aspect of the simulation is that it is not performed in the transient mode. In the measurement method of [10], it is assumed that enough time is elapsed for each step to achieve steady-state. As a result, non-transient numerical methods can be utilized in the simulations. Hence, much faster steady-state results can be achieved in PSM.

In order to extract valuable information from a single simulation, various probes should be incorporated. A charge concentration integrating probe for the PPD electron population should be used which is indicated with a dotted rectangle in Fig. 1 (b). Moreover, in three points with highest potentials inside the PPD, three probes to extract the PPD potentials and three probes for the electron quasi-fermi level (eQFL) potentials can be used. Also, two probes should be utilized to extract the potential and the eQFL potential of the FD for FD operational analysis. The potential and eQFL probes are indicated with white crosses inside PPD and FD. The probes are named A, B, C and F for future usage.

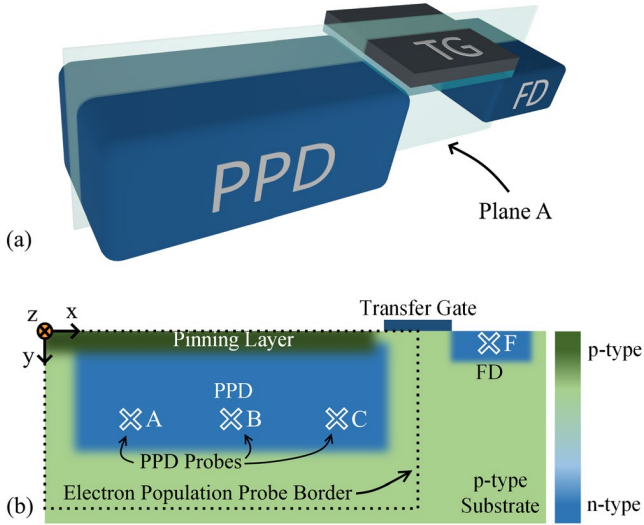

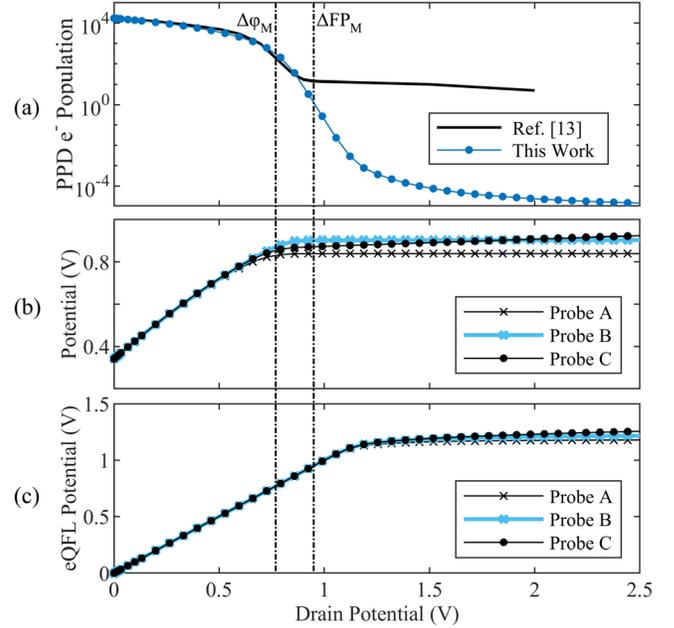

Fig. 1. (a) A 3-D PPD CIS representation including PPD, FD and TG. (b) A detailed device representation in plane A. The location of the potential and eQFL probes are indicated by crosses.

As mentioned earlier, in order to perform the simulations, transfer and reset gates are turned ON and the reset drain potential is swept from 0 to 3.3 V. In order to verify the method and to make sure that the simulation software is capable of providing accurate results, a conventional device performance reported in [11] and analyzed in [13] is verified. The models used in the simulations are Shockley-Read-Hall recombination model, concentration dependent mobility and field dependent mobility models. The dimensions and the dopant concentrations are chosen close to the values provided in [11, 13] and the results are illustrated in Fig. 2. As can be seen in Fig. 2, the pinning simulation electron population is close to the reported values in [11, 13].

As the drain potential increases, the FD potential directly follows. Different points inside the PPD follow the drain potential through the FD node. The eQFL potential of the PPD also follows the drain eQFL potential through the FD node.

In the beginning of the drain potential increment, the PPD electron population is a semi-linear function of the drain potential [11, 12, 16]. At a certain level of the drain potential, the PPD potential stops following the drain potential. Above the mentioned drain potential level which is called the maximum potential variation ($\Delta\varphi_M$), the PPD electron population becomes an exponential function of the drain potential [11, 12, 16].

In potential levels above $\Delta\varphi_M$, the PPD eQFL still follows the FD eQFL until PPD becomes empty of charges [11, 12, 16]. The maximum eQFL potential variation ($\Delta FP_M$) is defined as the drain potential level where the electron population reaches a single electron [12, 16]. The obtained values of $\Delta\varphi_M$ and $\Delta FP_M$ are close to the measured and simulated results reported in [11, 13].

Fig. 2. (a) PPD electron population for experiment-based results of [13] and PSM results of the present work. The difference between the curves at high drain potentials is due to charge partition phenomenon, not modeled in PSM. (b) PPD potentials and (c) eQFL potentials as functions of reset drain potential.

In practice, PPD electron population is limited by the charge partition phenomenon and is usually above a single electron [11, 13, 16]. As can be seen in Fig. 2, in PSM simulations which are not affected by charge partition, the minimum electron population can be less than one electron; the minimum electron population value depends on the PPD and TG design. Moreover, the eQFL potential variation can be more than the definition of $\Delta FP_M$.

The advantage of the pinning simulation method is in showing $\Delta\varphi_M$ and $\Delta FP_M$ in different locations of the PPD with only a single simulation. In the present manuscript, the proposed method is used as a fast PPD analysis tool.

In every design, in order to make sure that PPD is certainly pinned, the minimum PPD electron population should be less than a single electron. Moreover, the values of $\Delta\varphi_M$ and $\Delta FP_M$ affect the variation of electron population and should also be considered in design process.

## III. NEW PPD STRUCTURES

With respect to the results of section II, the goal of the new PPD structure design is to create steps of pinning potentials with the lowest pinning potential at point A and the highest pinning potential at point C. With such a structure which is named multi-level pinned photodiode (MPPD), the last remaining electrons would be near TG, available for fast transfer to FD.

The pinning process described in [12, 16, 17] indicates that in most of the planar rectangular PPDs with a typical cross-section similar to Fig. 1 (b), the pinning process occurs in the PPD y-direction; the pn junctions at the surface and in the depth of the PPD are responsible for the pinning process. One way to change the pinning potential throughout the PPD and design an MPPD is to change the PPD impurity dopant concentration which requires modifications regarding the ion implantation steps [9].

In the first suggestion of the present work, the pinning process of the PPD is modified by using lines of impurity dopants instead of a rectangular structure. The suggested PPD mask is illustrated in Fig. 3 (b).

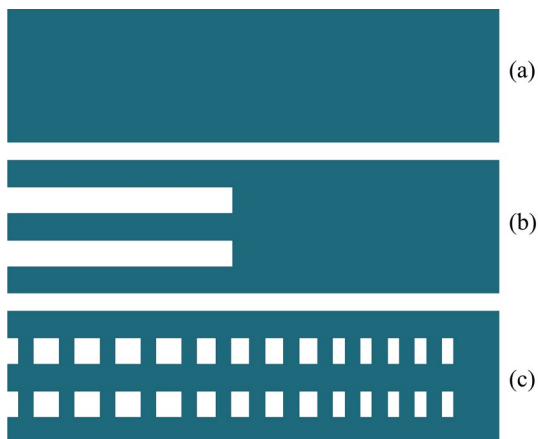

Fig. 3. (a) The rectangular PPD mask which is 2.8 $\mu$m×0.8 $\mu$m. (b) The lines mask. (c) The holey mask.

The empty lines create some locations with the opposite dopant type in the PPD which considerably lower the pinning potential in their nearby regions. The lines region should be located far from TG to create the desired electric field.

The holey mask, Fig. 3 (c), is based on the same idea as the lines mask but designed to create more pinning steps and as a result, smoother electric field; desired in pushing the electrons towards TG. The holes that are included in the holey mask cause the pinning potential variations. Wide holes result in lower pinning potentials compared to tight holes since they introduce a wider volume of p-type regions in n-type PPD centroid. Wider p-type regions, pin down the PPD potential more effectively and cause lower pinning potentials. As a result, pinning potential steps is created without the need of additional masks or process steps.

## IV. RESULTS AND DISCUSSION

The pinning simulation is used to study the pinning potential variation inside the pixel and also to observe the pinning effect on the PPD electron population. PPDs are simulated with the parameters and simulation conditions similar to the ones used to obtain Fig. 2. The spatial dimensions and the implantation masks used to obtain Fig. 2 are modified in accordance with the proposed masks in Fig. 3.

Moreover, the introduction of holes and lines in the masks lowers the equilibrium full-well capacity (EFWC). Hence, the PPD dopant concentrations of the lines and the holey PPDs are multiplied by appropriate values to result in equal EFWCs for pixels using masks of Fig. 3.

The PSM results related to the rectangular mask are illustrated in Fig. 4. As can be seen in Fig. 4, in a rectangular pixel, point B's potential is comparable to points A and C which causes the last remaining electrons to be spread out in the pixel. In the pixels with large PPDs where point B is far from TG and also in high speed applications, point C should have a considerably higher potential in order to collect the last remaining electrons near TG. In the rectangular pixel, a considerable number of electrons should diffuse to point C in order to emit to FD.

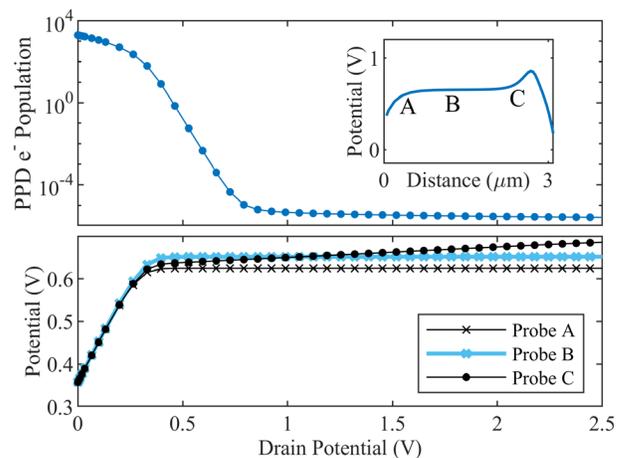

Fig. 4. The results of pinning simulation method for the pixel with the rectangular mask. The inset shows the potential profile in the PPD along probes A, B and C with a drain potential of 3.3 V.

The PSM results related to the lines mask are illustrated in Fig. 5.

In Fig. 5, probe A's maximum or pinning potential ($\Delta\varphi_{MA}$) is close to zero and fixed because of the opposite-type impurity dopants nearby. A very small $\Delta\varphi_M$ can be translated as a very low electron concentration. As expected, the lines cause a large step of pinning potential and point C has almost the highest potential. However, the results indicate that electrons are not distributed in a desired manner and very few electrons can be collected near probe A. On the other side of the pinning step, large number of electrons are collected in the right half of PPD.

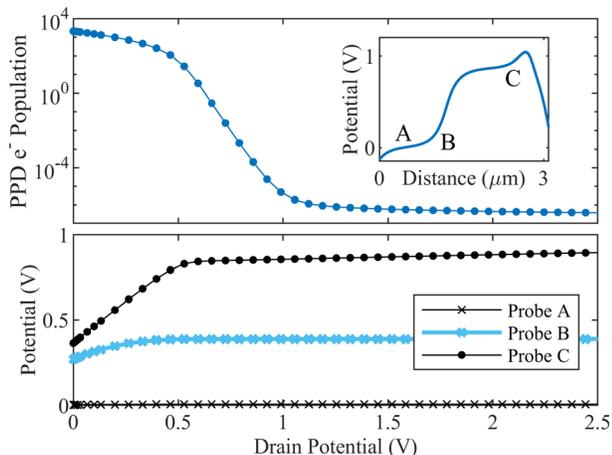

Fig. 5. The results of pinning simulation method for pixel with the lines mask. The inset shows the potential profile in the PPD along probes A, B and C with a drain potential of 3.3 V. Multi-level pinning caused by lines of dopants can be observed as the difference between maximum probes potentials.

In order to engineer PPD electric field in a desired manner, the holey mask is proposed in Fig. 3 (c) with 80 nm minimum feature size. The holey mask introduces smaller steps of dopant variations, expected to cause smaller pinning steps. The PSM results regarding the holey mask are illustrated in Fig. 6.

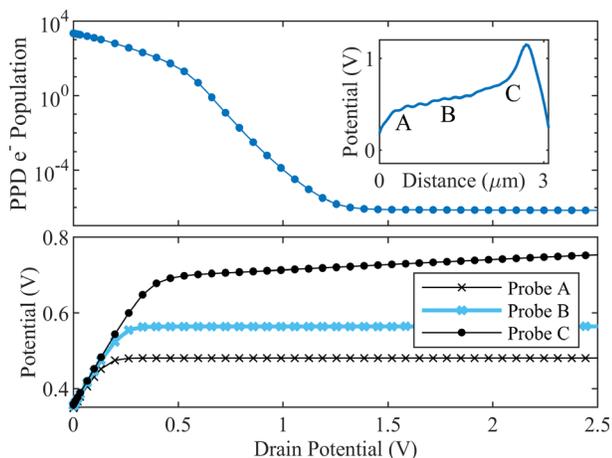

Fig. 6. The results of pinning simulation method for the pixel with the holey mask. The inset shows the potential profile in the PPD along probes A, B and C with a drain potential of 3.3 V.

Pinning potentials of the probes in Fig. 6 create a fair distribution of charges inside PPD and at the same time create a considerable electric field to push the electrons towards TG. Moreover, the mask design ensures an increase in the number of pinning steps which can be translated as smoother electric field extended throughout the PPD.

With the holey PPD, the need for electrons to diffuse inside PPD is minimized. In order to further investigate the pixels, similar time-based simulations are performed for the rectangular and holey pixels. The time-based simulations are performed with the standard 4-transistor PPD pixel timing, considering a 300 ns charge transfer time. The time-based results during the first 100 ns of the transfer phase are illustrated in Fig. 7.

In order to quantify the results, the transfer time is defined as the time needed for the electron population to reach from 90% to 10% of the maximum value. With the mentioned definition, the use of holey mask lowers the charge transfer time by 31%.

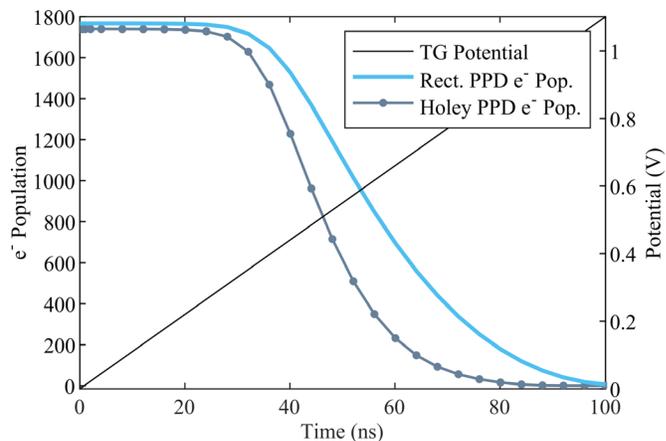

Fig. 7. The time-based results during charge transfer for the pixels with the rectangular and holey masks.

In the following paragraphs, the logic behind faster charge transfer with the holey mask is investigated; which is concluded to be the lowered electron diffusion time. Upon electrons collection in the vicinity of transfer gate, the diffusion time is minimized. In a rectangular pixel where PPD electric field engineering is not performed, electrons are collected starting from the middle of PPD. Hence, the average diffusion distance for the PPD with the rectangular mask would be 0.7 µm (half of the distance from the middle of the PPD to TG). The diffusion time for the rectangular PPD can be approximately calculated by [18, 19]:

$$\Delta t_d = \frac{L^2}{2D_n} = \frac{\left(0.7 \times 10^{-4} \text{ cm}\right)^2}{2 \times 12 \text{ cm}^2/s} = 0.2 \text{ ns} \tag{1}$$

where $D_n$ is the electron diffusion coefficient. The difference between the rectangular and holey PPDs can be observed by comparing the average diffusion time and the average thermionic emission time needed for the transport of an electron.

With respect to the time-based results provided in [17], thermionic emission time to transfer 20000 e- is less than 1000 ns in a PPD with a well-designed transfer gate. This results in an average 0.05 ns thermionic emission time for each electron in the pixel of [17]. As can be seen in Fig. 7, the present work's rectangular and holey pixels have a charge transfer time, comparable to the pixel of [17]. This results in an average thermionic emission time which is approximately a quarter of the average diffusion time. Hence, the diffusion time is not negligible compared to thermionic emission time, even for a PPD with a 2.8 µm length and can be considered as the reason for faster charge transfer in the holey PPD compared to the rectangular PPD.

## V. CONCLUSION

In order to design and analyze a PPD pixel with multiple pinning potentials, a fast simulation approach is presented. The pinning simulation method which is a modified version of a well-known measurement method, is capable of extracting valuable information from a 3-D device in a single run. The simulation approach is used to analyze a rectangular PPD which is not suitable for fast or large pixel designs. The problem is discussed with focus on pinning process and a suggestion is presented to overcome the problem; the holey mask. The simulation results indicate that the transfer time is lowered by 31% without any added fabrication process steps. Finally, the logic behind faster charge transfer in the pixel with holey PPD is concluded to be reduced electron diffusion time.